\documentclass[conference]{IEEEtran}
\IEEEoverridecommandlockouts
\usepackage{cite}
\usepackage{amsmath,amssymb,amsfonts}
\usepackage{algorithmic}
\usepackage{graphicx}
\usepackage{textcomp}
\usepackage{xcolor}
\usepackage{subcaption}
\usepackage{tabularx}
\usepackage{multirow}
\def\BibTeX{{\rm B\kern-.05em{\sc i\kern-.025em b}\kern-.08em
    T\kern-.1667em\lower.7ex\hbox{E}\kern-.125emX}}
\newcommand{\myauthorrefmark}[1]{\textsuperscript{#1}}
\begin{document}

\title{PermutEx: Feature-Extraction-Based Permutation — A New Diffusion Scheme for Image Encryption Algorithms\\
}

\author{
\IEEEauthorblockN{
    Muhammad Shahbaz Khan\myauthorrefmark{1},
    Jawad Ahmad\myauthorrefmark{1},
    Ahmed Al-Dubai\myauthorrefmark{1},
    Zakwan Jaroucheh\myauthorrefmark{1}
}
\IEEEauthorblockN{
    Nikolaos Pitropakis\myauthorrefmark{1},
    William J. Buchanan\myauthorrefmark{1},
}
\IEEEauthorblockA{\myauthorrefmark{1}\textit{School of Computing, Engineering and the Built Environment}, \\
\textit{Edinburgh Napier University}, \\
Edinburgh, UK, \\
Emails: \{muhammadshahbaz.khan, j.ahmad, a.al-dubai, z.jaroucheh, n.pitropakis, b.buchanan\}@ napier.ac.uk}
}

\maketitle

\begin{abstract}
Traditional permutation schemes mostly focus on random scrambling of pixels, often neglecting the intrinsic image information that could enhance diffusion in image encryption algorithms. This paper introduces PermutEx, a feature-extraction-based permutation method that  utilizes inherent image features to scramble pixels effectively. Unlike random permutation schemes, PermutEx extracts the spatial frequency and local contrast features of the image and ranks each pixel based on this information, identifying which pixels are more important or information-rich based on texture and edge information. In addition, a unique permutation key is generated using the Logistic-Sine Map based on chaotic behavior. The ranked pixels are permuted in conjunction with this unique key, effectively permuting the original image into a scrambled version. Experimental results indicate that the proposed method effectively disrupts the correlation in information-rich areas within the image resulting in a correlation value of 0.000062. The effective scrambling of pixels, resulting in nearly zero correlation, makes this method suitable to be used as diffusion in image encryption algorithms.
\end{abstract}
\vspace{10pt}
\begin{IEEEkeywords}
Diffusion, permutation, feature extraction, spatial frequency, local contrast, Josephus permutation, chaos.
\end{IEEEkeywords}

\section{Introduction}
\begin{figure*}[!htb]
    \centering
    \begin{subfigure}[b]{0.3\textwidth}
        \includegraphics[width=\textwidth]{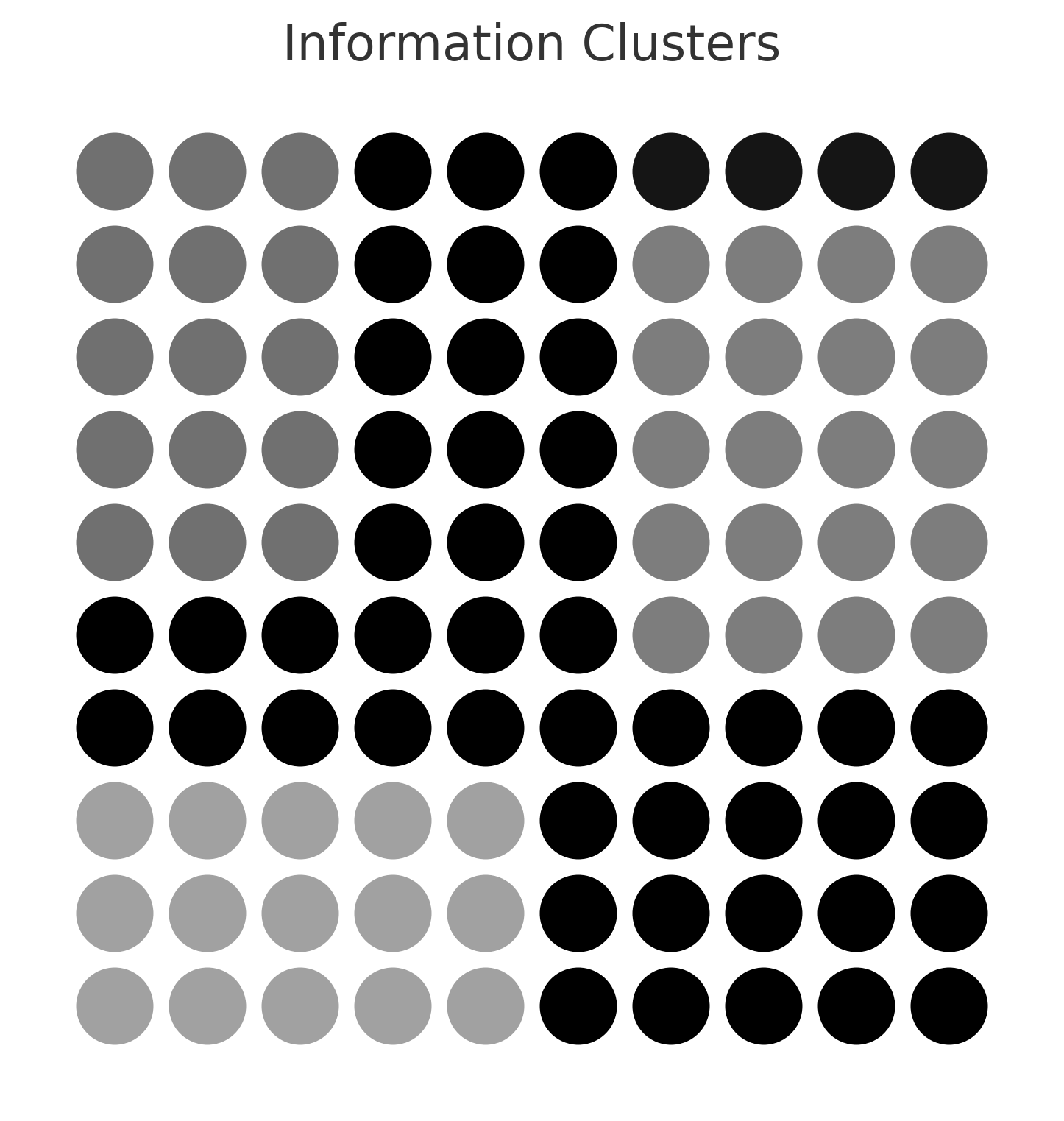}
        \caption{}
        \label{subfig:1-a}
    \end{subfigure}
    \hfill
    \begin{subfigure}[b]{0.3\textwidth}
        \includegraphics[width=\textwidth]{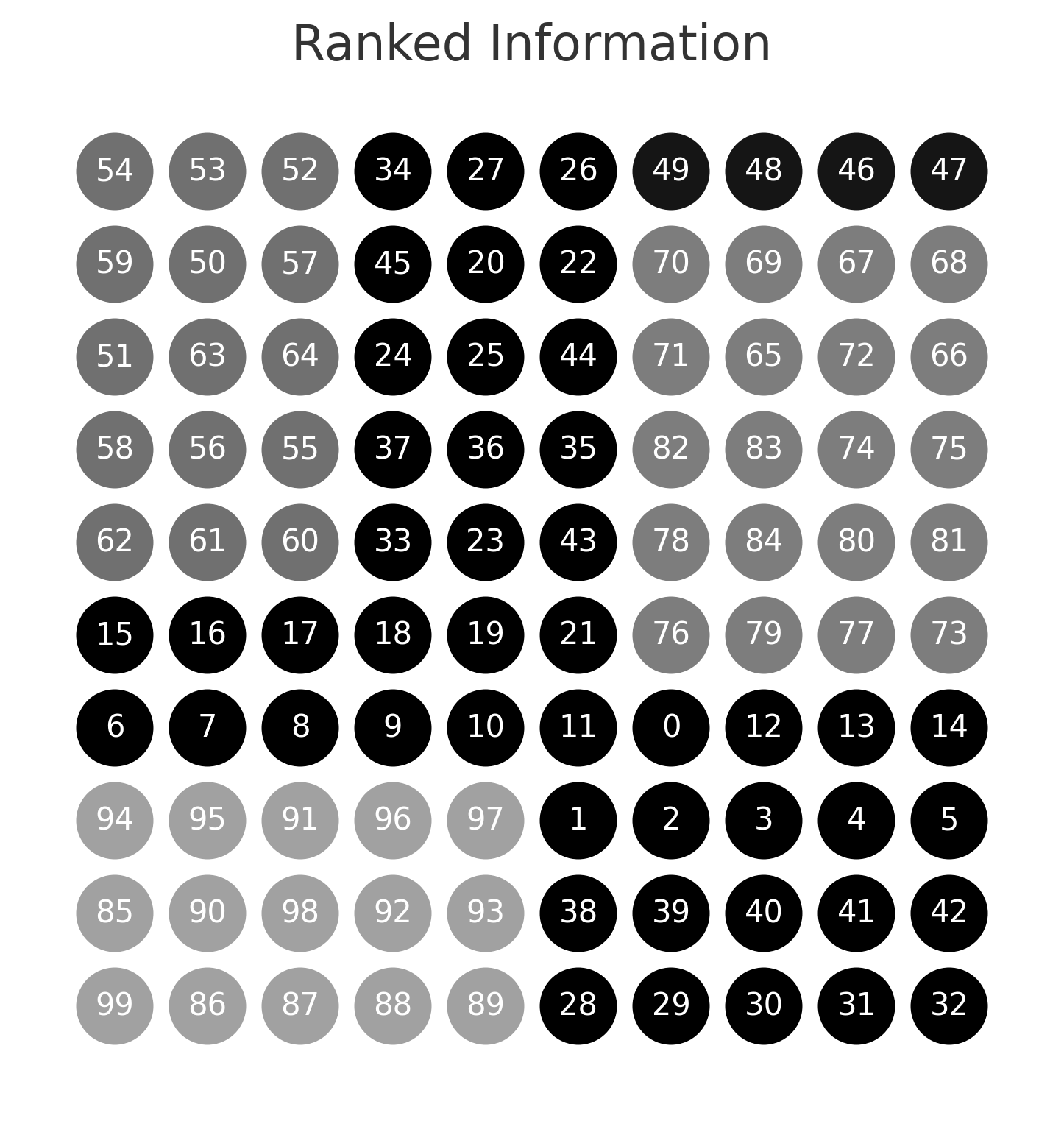}
        \caption{}
        \label{subfig:1-b}
    \end{subfigure}
    \hfill
    \begin{subfigure}[b]{0.3\textwidth}
        \includegraphics[width=\textwidth]{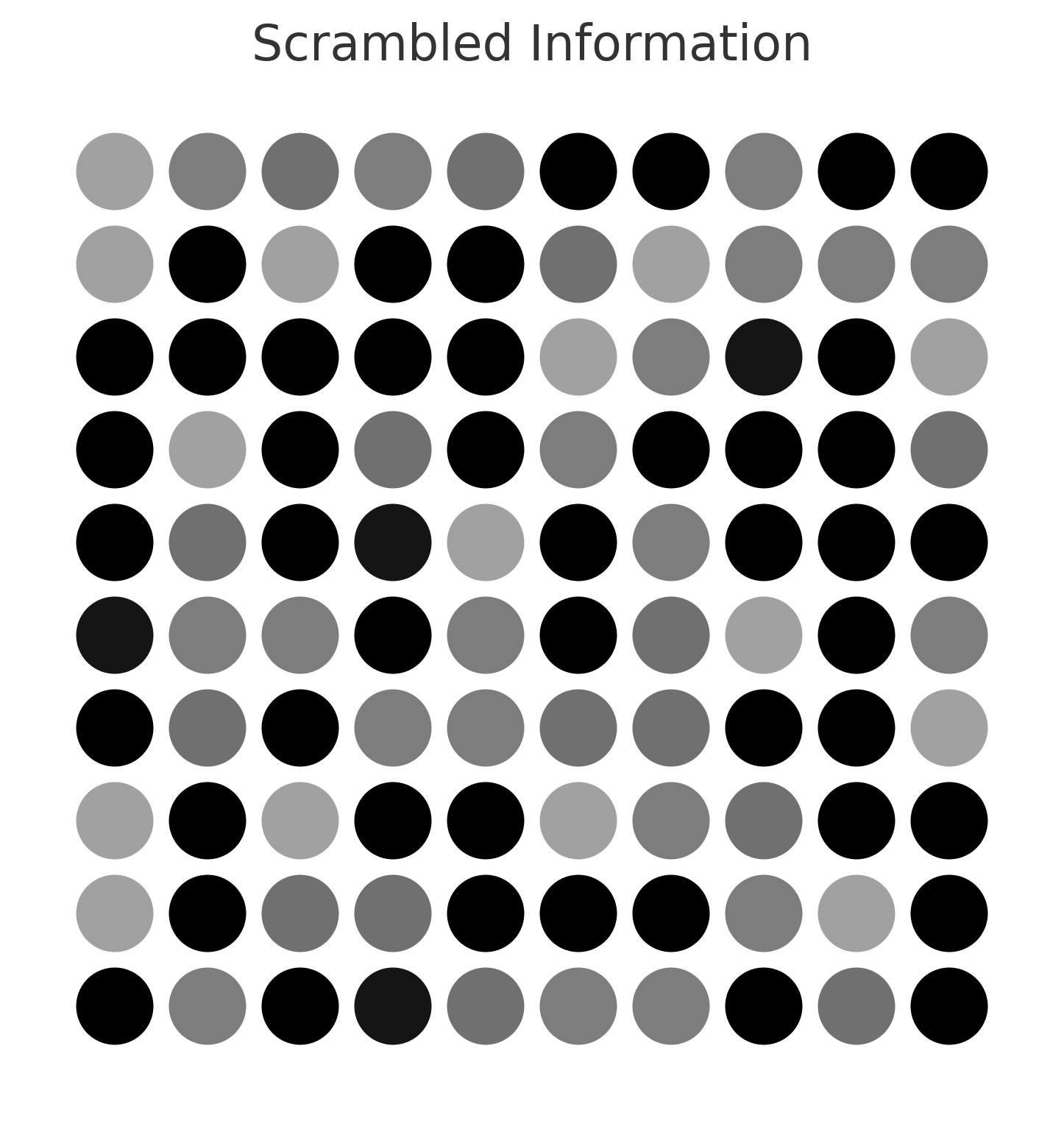}
        \caption{}
        \label{subfig:1-c}
    \end{subfigure}
    \caption{Basic workflow of information-based scrambling}
    \label{fig:basic_workflow}
\end{figure*}

In digital image security, image encryption plays a vital role in ensuring confidentiality and integrity \cite{ahmed_2022_a, sahu_2023_a, khan_2023_srss}. Image encryption involves two basic processes, i.e., confusion and diffusion. According to Claude Shannon \cite{shannon_1949_communication}, confusion refers to changing the values of the pixels based on a key and is usually achieved by substituting one value for another. Diffusion, on the other hand, refers to changing the position of the pixels based on a key. This is usually achieved through methods like the permutation. Traditional permutation schemes often focus on the random scrambling of pixels, neglecting intrinsic image information, which could be utilised for a more effective permutation strategy. Such schemes include chaotic row-column shuffling \cite{8653280, 9858333, 10.1007/978-3-030-01177-2_56}, classic Josephus permutation \cite{hua_2019_image}, and improved Josephus permutation \cite{gao_2006_a}. Although these methods may offer a certain level of unpredictability, they often fall short of exploiting the image's inherent features, which could be important for enhancing the quality of diffusion. To achieve effective scrambling, this paper argues that the permutation algorithms should utilize intrinsic image features.

Features like spatial frequency and local contrast both relate to the variation in intensity values of an image, but they focus on different aspects and scales of variation. Spatial frequency refers to the rate at which pixel intensities change in an image \cite{decesarei_2013_early, marr_1982_vision, devalois_1980_spatial}. It can be thought of as the fineness or coarseness of patterns in an image. High Spatial Frequency refers to rapid changes in pixel values over short distances. This is often seen in areas with detailed textures or sharp edges. For example, in an image of a zebra, the black and white stripes represent a high spatial frequency because of the rapid alternation of contrasting colors. Low Spatial Frequency, on the other hand, depicts slow changes in pixel values over larger distances. This would be seen in areas with gradual intensity changes, like a smooth gradient sky during sunset. Spatial frequency can be analyzed using Fourier transforms. Furthermore, local contrast or local dynamic range refers to the difference in intensity between a pixel and its immediate surroundings. It's a measure of how much a pixel stands out from its neighbors- \cite{zhou_2014_global, peli_1990_contrast}. A high local contrast means that a pixel that has a much different intensity than its neighbors. For instance, if a single white pixel is surrounded by black pixels, this white pixel has a high local contrast because of its stark difference from its immediate environment. 

This paper introduces PermutEx, a novel approach to permutation that takes into account the inherent features within the image for effective scrambling of pixels. Unlike traditional methods, our scheme examines the spatial frequency and local contrast of each pixel, ranking them based on this information. This information is used to scramble the pixels, disrupting the correlation in information-rich areas within the image effectively. Fig. \ref{fig:basic_workflow} depicts the basic workflow of the proposed permutation scheme. The clusters in Fig. \ref{subfig:1-a} represent the information in an image such as edges, shapes, etc. The darker dots represent the information-rich areas, and the lighter dots represent the information-poor areas. Fig. \ref{subfig:1-b} depicts the pixels ranked on the basis of information content. Each dot is annotated with its rank where dots with higher ranks represent higher information content and in Fig. \ref{subfig:1-c}, the dots have been scrambled based on the information ranking.

The main contributions of this paper are:

\begin{enumerate}
    \item A novel feature-extraction or image-information-based permutation scheme, PermutEx is proposed that incorporates feature extraction to enhance pixel scrambling. 
    \item The intrinsic image features, such as spatial frequency and local contrast are employed for effective pixel rearrangement. Pixels are ranked on the basis of the maximum and minimum information and are scrambled accordingly.
    \item A unique permutation key is generated using hybrid Logistic-Sine Map having improved chaotic behaviour and large chaotic range.This key is used in conjunction with the pixel ranking criteria to scramble the pixels and break the correlation effectively.
\end{enumerate}

\section{The Proposed Permutation Scheme---PermutEx}
The complete steps involved in the proposed PermutEx scheme, depicted in Fig.\ref{fig:fig2-Main_technique}, are as follows:\\

\begin{itemize}
    \item \textbf{Step 1. Read Plaintext Image:} Read the grayscale \(256 \times 256\) baboon image as the plaintext image matrix \(I\) represented as (1).
    \begin{equation}
        \
I = \{ I(x, y) \}_{(x=1, y=1)}^{256 \times 256}
\
    \end{equation}

\item \textbf{Step2. Calculate Spatial Frequencies:} To extract the information on how the pixel intensity values in the image vary over space, the spatial frequency information has been calculated using Fast Fourier Transform (FFT). High frequencies represent the regions with more information/complexity, such as the edges, textures, and other fine-grained details in the image. The low frequencies, on the other hand, represent the basic structures and shapes in the image. In Fig.\ref{fig:fig3-b}, the brighter spots indicate where the maximum information is, while the low frequency components depicting basic shape/structure information are visualised in Fig. \ref{fig:fig3-c}. The FFT of the plaintext image \(I(x,y)\) is calculated by:

\begin{equation}
F(u,v) = \sum_{x=0}^{256-1} \sum_{y=0}^{256-1} I(x,y) \cdot e^{-j2\pi \left( \frac{ux}{256} + \frac{vy}{256} \right)}
\end{equation}

\begin{equation}
F(u,v) = [F_1, F_2, F_3, \ldots, F_{256}]
\end{equation}

Where \(F(u,v)\) is the Fourier transform of \(I(x,y)\), \(256 \times 256\) shows the dimension of the image, \(j\) is the imaginary unit, and \([F_1, F_2, F_3, \ldots, F_{256}]\) are complex numbers.

The zero-frequency component is moved to the centre of the array as:

\begin{equation}
F_{\text{shifted}}(u,v) = F\left(u-\frac{256}{2}, v-\frac{256}{2}\right)
\end{equation}

The magnitude of the complex numbers in \(F_{\text{shifted}}\) is calculated as:

\begin{equation}
F_{\text{mag}}(u,v) = \sqrt{\text{Re}(F_{\text{shifted}})^2 + \text{Im}(F_{\text{shifted}})^2}
\end{equation}

To enhance the visibility of the components, the logarithmic scaling is applied:

\begin{equation}
F_{\text{mag\_log}}(u,v) = \log(F_{\text{mag}}(u,v) + 1)
\end{equation}

The final step is to normalize the logarithmically scaled magnitude to fit within the grayscale range \([0,1]\):

\begin{equation}
F_{\text{norm}}(u,v) = \frac{F_{\text{mag\_log}}(u,v) - \min(F_{\text{mag\_log}})}{\max(F_{\text{mag\_log}}) - \min(F_{\text{mag\_log}})}
\end{equation}\\

\item \textbf{Step 3. Calculate Local Contrast:} The local contrast is calculated by using a local window around each pixel and calculating the contrast within that window. The contrast is often defined as the standard deviation of the pixel values within the window and essentially captures the local variability or range of pixel values, which can be thought of as a very rudimentary form of "texture" or "detail" at that pixel location. Visually, high contrast values represent the areas where the image changes dramatically over a small distance, like the edges of objects or textures as shown in Fig. \ref{fig:fig3-d} and the low contrast values represent smooth, slowly varying areas, which can also be visualised in Fig. \ref{fig:fig3-d}.

\begin{figure}[t]
    \centering
    \includegraphics[width=0.99\linewidth]{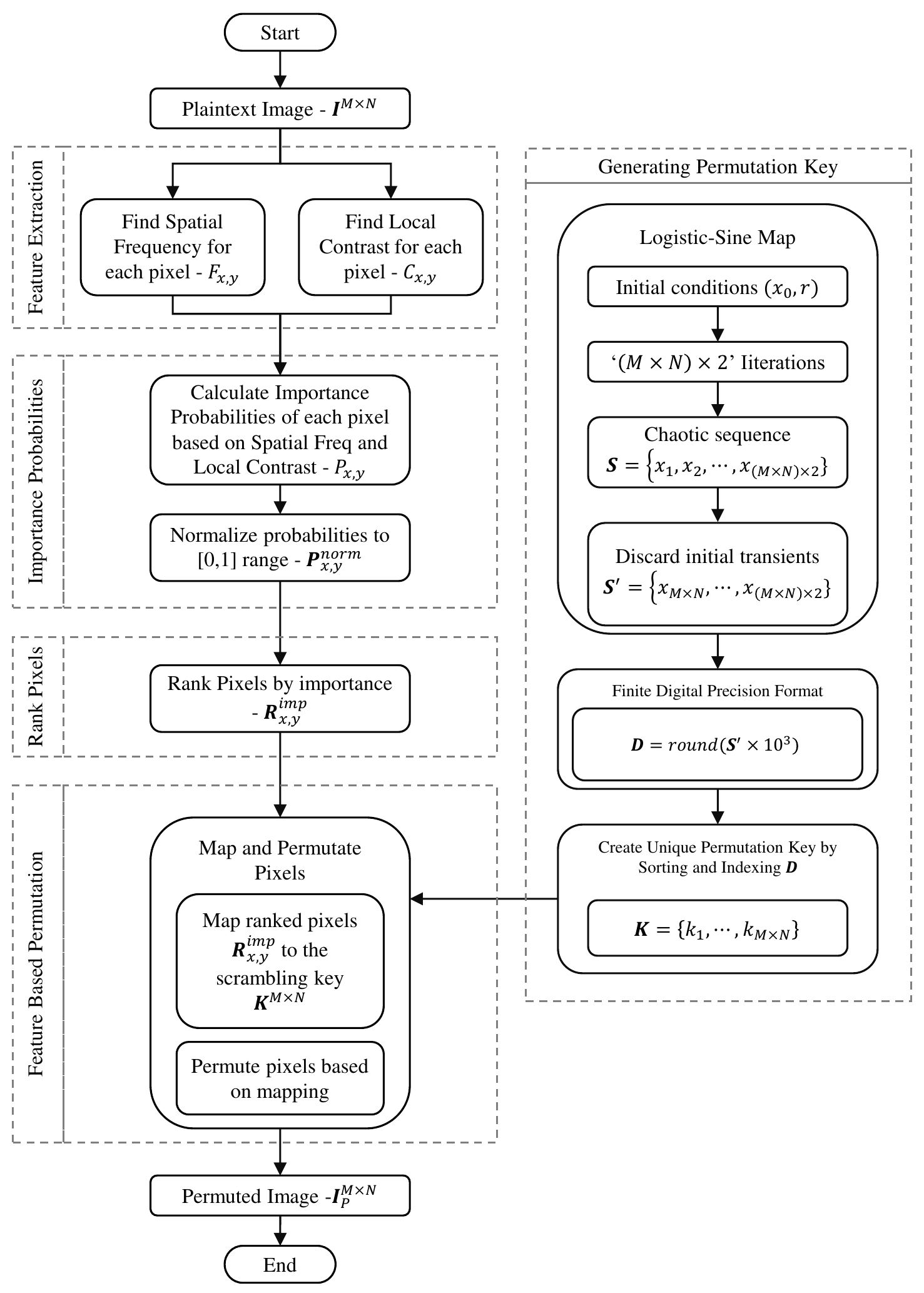}
    \caption{PermutEx –The proposed feature-extraction-based Permutation Scheme.}
    \label{fig:fig2-Main_technique}
\end{figure}

For our \(256 \times 256\) input image \(I\), where \(I(x,y)\) represents the intensity of the pixel at location \( (x,y) \), the local contrast is calculated by choosing a local window \(W\) of size \(m \times m\) centered at location \( (x,y) \). We selected a \(3 \times 3\) window and it is represented as:
\begin{align}
W(x,y) = \{ & I(i,j) \mid i \in [x-(m-1)/2, x+(m-1)/2], \notag \\
& j \in [y-(m-1)/2, y+(m-1)/2] \}
\end{align}

Then calculate the mean \( \mu \) of the pixel intensities within \( W \):

\begin{equation}
\mu = \frac{1}{m^2} \sum_{(i,j) \in W} I(i,j)
\end{equation}

Then calculate the standard deviation \( \sigma \) of the pixel intensities within \( W \), which serves as the local contrast \( C(x,y) \) expressed as (10) and visually depicted as shown in Fig. \ref{fig:fig3-d}.

\begin{equation}
C(x,y) = \sigma = \sqrt{\frac{1}{m^2} \sum_{(i,j) \in W} (I(i,j) - \mu)^2 }
\end{equation}\\

\item \textbf{Step 4. Calculate Importance Probabilities:} Information probabilities are calculated by taking the average of corresponding values from the Spatial Frequency array \( F(u,v) \) having frequency values for each pixel as \( F_{x,y} \) and the Local Contrast Array \( C(x,y) \) having contrast values for each pixel as \( C_{x,y} \). This new array captures a blended form of both spatial frequency and local contrast information. The probability \( P \) of a pixel at location \( (x,y) \) is calculated as:

\begin{equation}
P_{x,y} = \frac{F_{x,y} + C_{x,y}}{2}
\end{equation}

\begin{figure}[!b]
    \centering
    \begin{subfigure}{0.45\linewidth}
        \includegraphics[width=\linewidth]{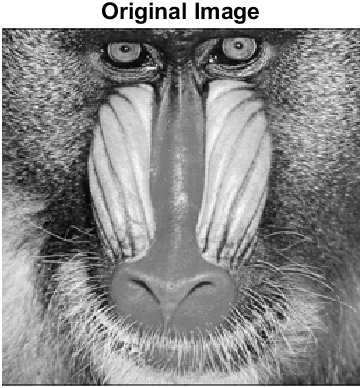}
        \caption{}
        \label{fig:fig3-a}
    \end{subfigure}
    \hfill
    \begin{subfigure}{0.45\linewidth}
        \includegraphics[width=\linewidth]{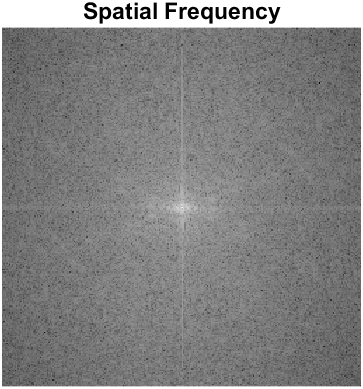}
        \caption{}
        \label{fig:fig3-b}
    \end{subfigure}
    \begin{subfigure}{0.46\linewidth}
        \includegraphics[width=\linewidth]{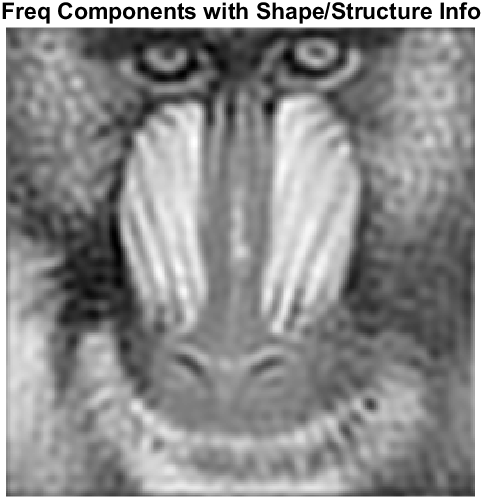}
        \caption{}
        \label{fig:fig3-c}
    \end{subfigure}
    \hfill
    \begin{subfigure}{0.45\linewidth}
        \includegraphics[width=\linewidth]{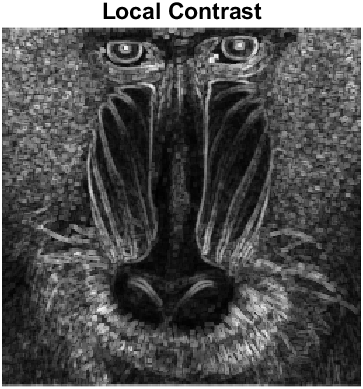}
        \caption{}
        \label{fig:fig3-d}
    \end{subfigure}
    \caption{Visual representation of extracted features: (a) original image, (b) spatial frequency, (c) frequency components with shape/structure info, and (d) local contrast.}
    \label{fig:fig3-features}
\end{figure}

This metric provides a measure of "importance" or "information richness" at each pixel location. Pixels with higher values in the probabilities array would mean that both their spatial frequency and local contrast are high, making them significant in terms of edges, texture, and overall complexity. Whereas, the pixels with lower values are less significant in terms of both spatial frequency and local contrast. These are probably smoother areas with less detail.\\

\item \textbf{Step 5. Normalize Importance Probabilities:} Normalizing the "probabilities" array to a range of \(0\) to \(1\) is done using the Min-Max normalization method. First, calculate the minimum \(\text{minProb}\) and maximum \(\text{maxProb}\) values of the "probabilities" array.

\begin{equation}
\text{minProb} = \min_{x,y}(P_{x,y})
\end{equation}

\begin{equation}
\text{maxProb} = \max_{x,y}(P_{x,y})
\end{equation}

The probabilities are then normalized to the range \([0,1]\) using the formula:

\begin{equation}
P_{x,y}^{\text{norm}} = \frac{P_{x,y} - \text{minProb}}{\text{maxProb} - \text{minProb}}
\end{equation}\\

\item \textbf{Step 6. Rank Pixels by Importance:} In this step, the grayscale pixels are ranked by their importance based on their normalized probabilities. The sorting is done in descending order, meaning the most important (highest probability) pixels come first. The pixels ranked on the basis of importance are visualized in Fig. \ref{fig:fig4-ranked_pixels}(b), where yellow edges represent high values of probabilities depicting the areas having a lot of information according to the probability metric (a combination of spatial frequency and local contrast). These are the areas of the image with higher texture or edge information. On the other hand, the blue background represents the areas having less information according to the probability metric. These are likely to be smoother, less textured areas of the image.

The ranking is done as follows:

\begin{enumerate}
    \item The normalized probability matrix \( P^{\text{norm}} (256 \times 256) \) is flattened into a vector \( p' \) of length \( 256 \times 256 \).
    \begin{equation}
    p' = \text{Flatten}(P^{\text{norm}})
    \end{equation}
    
    \item The vector \( p' \) is then sorted in descending order as follows:
    \begin{equation}
    p_{\text{sorted}} = \text{Sort}(p', \text{'descend'})
    \end{equation}
    
    \item The original indices corresponding to the sorted vector \( p_{\text{sorted}} \) are stored in \( R^{\text{imp}} \). These indices indicate the positions of the pixels in the sorted vector \( p' \).
    \begin{equation}
    R^{\text{imp}} = \text{Indices of } p_{\text{sorted}} \text{ in } p'
    \end{equation}
\end{enumerate}

The \( R^{\text{imp}} \) array contains the indices of the pixels in the sorted normalized probability matrix, sorted by their importance (highest probability first). We use these indices to prioritize or select pixels for further processing, based on their calculated importance.\\

\begin{figure}[!ht]
    \centering
    \begin{subfigure}{\linewidth}
        \centering
        \includegraphics[width=0.8\linewidth]{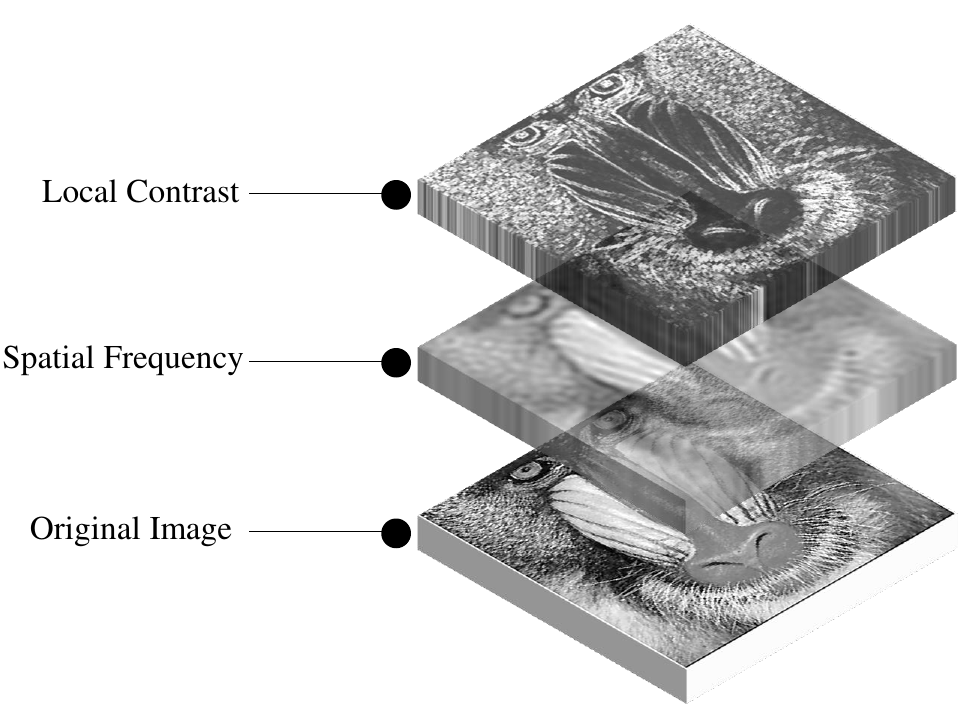}
        \caption*{(a)}
        \label{fig:fig4-a}
    \end{subfigure}
    
    \vspace{1em}
    
    \begin{subfigure}{\linewidth}
        \centering
        \includegraphics[width=0.8\linewidth]{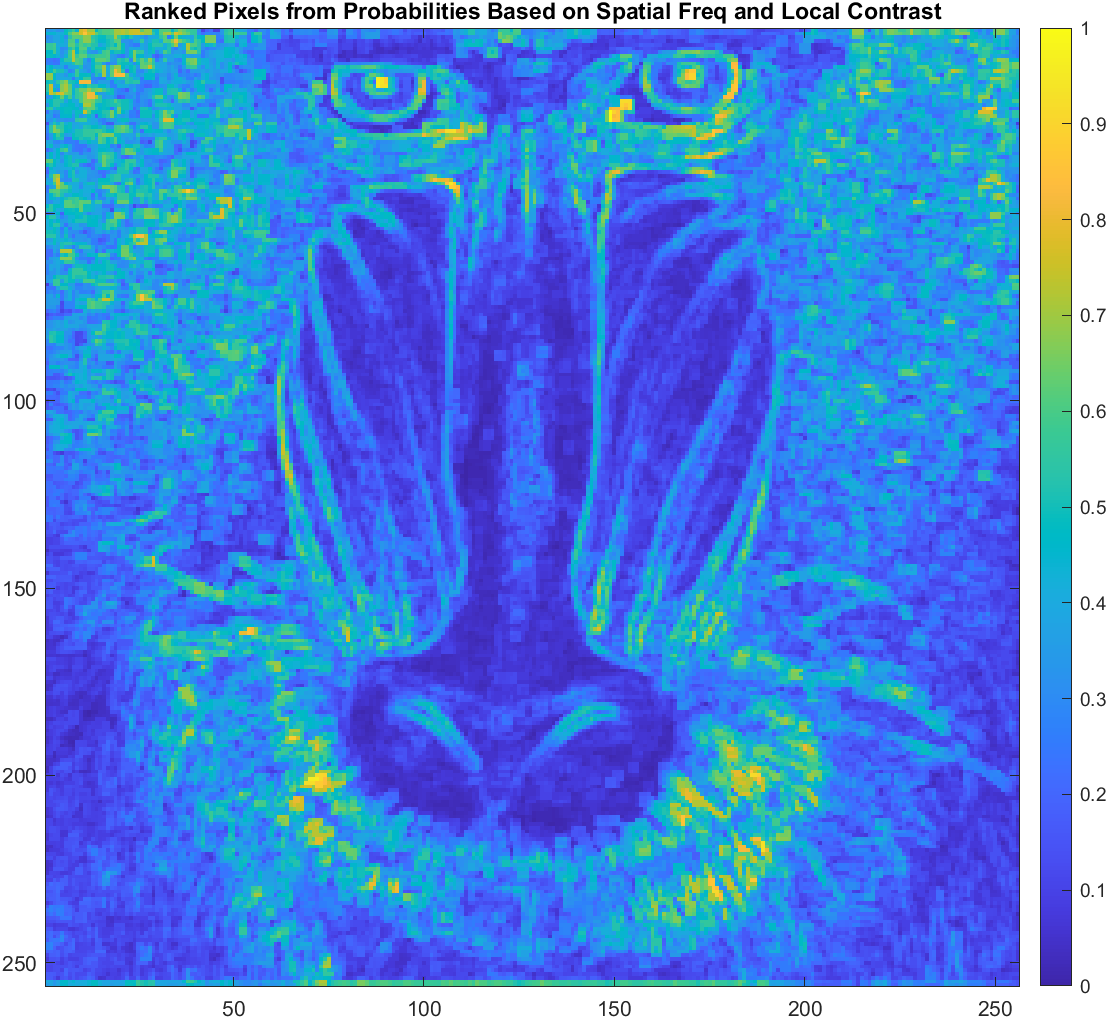}
        \caption*{(b)}
        \label{fig:fig4-b}
    \end{subfigure}
    
    \caption{Visualization of ranked pixels based on importance: (a) Information acquisition from spatial frequencies and local contrast, (b) Visualization of significant and less important pixels.}
    \label{fig:fig4-ranked_pixels}
\end{figure}

\item \textbf{Step 7. Generate the Permutation Key:} A permutation key \( K \) is generated using the Logistic Sine Map. This is a chaotic map used for generating a sequence of values that appear highly random. The highly chaotic permutation key is depicted in Fig. \ref{fig:fig5-permutation_results}(c) and is generated using the following steps:

\begin{enumerate}
    \item The number of iterations is set twice the number of pixels in the input image \( I \).
    \begin{equation}
    \text{Iterations} = 2 \times \text{numel}(I)
    \end{equation}

    \item The hybrid logistic sine map is defined by:
    \begin{align}
    X_{N+1} = \left( L(r, X_N) + \right. \nonumber \\
    \left. S((4-r), X_N) \right) \mod 1
    \end{align}

    Where \( L(r, X_N) \) is the logistic map function and \( S((4-r), X_N) \) is the Sine Map function. Specifically, these are defined as:\\

    The Logistic Map: \( L(r, X_N) = r \times X_N \times (1 - X_N) \).\\
    The Sine Map: \( S((4-r), X_N) = \frac{(4-r) \times \sin(\pi \times X_N)}{4} \).\\

By combining these two, the hybrid map becomes:
\begin{align}
X_{N+1} &= \Bigg( r \times X_N \times (1-X_N) \nonumber \\
& + \frac{(4-r) \times \sin(\pi \times X_N)}{4} \Bigg) \mod 1
\end{align}

    \item This generates a chaotic sequence \( S \). To avoid the initial transients that may not represent the chaotic behavior well, only the latter half of the sequence is taken and the new chaotic sequence \( S' \) is obtained.
    \begin{equation}
    S' = S[\text{end - numel}(I) + 1 : \text{end}]
    \end{equation}

    \item The values are scaled and rounded to integers to create a finite digital precision format sequence \( D \).
    \begin{equation}
    D = \text{round}(S' \times 10^3)
    \end{equation}

    \item The integer values from \( D \) are sorted and their original indices are stored in \textit{permuteSequence}. Then, the indices of this sorted array are again sorted to create the unique permutation key \( K \).

    \begin{equation}
        \text{permuteSequence} = \text{SortIndices}(D)
    \end{equation}
    \begin{equation}
        K=\text{SortIndices}(\text{permuteSequence})
    \end{equation}
\end{enumerate}

This generates a unique permutation key \( K \) which is used to map and permute the pixels in the next step.\\

\item \textbf{Step 8. Map and Permute to Get Scrambled Image:} In this step, we map the ranked list of original indices stored in \( R^{\text{imp}} \) to the generated permutation key \( K \). We then use this mapped list to rearrange the pixels in the original image, effectively scrambling it. From the previous steps, we have two sequences:
\begin{itemize}
    \item \( R^{\text{imp}} = [r_1, r_2, \ldots, r_{256 \times 256}] \) --- This is the sequence of original pixel indices sorted by their ranked importance.
    \item \( K = [k_1, k_2, \ldots, k_{256 \times 256}] \) --- This is the unique Permutation Key.
\end{itemize}

To create the final permuted indices to get the permuted image, we rearrange \( R^{\text{imp}} \) according to the key \( K \). The equation for this operation is:
\begin{equation}
    \text{permutedIndices}[z] = R^{\text{imp}}[r_z]
\end{equation}
For \( z = 1, 2, \ldots, 256 \times 256 \).

Here, \(\text{permutedIndices}[z]\) is the \( z^{\text{th}} \) element in the final permuted sequence, and \( R^{\text{imp}}[r_z] \) is the \( r_z^{\text{th}} \) element in the ranked sequence.

This is how the ranked pixels are scrambled according to the unique permutation key, resulting in a permuted or scrambled image.

\end{itemize}

\section{Results and Analysis}
The proposed permutation scheme uses the intrinsic information of the plaintext image to scramble the images in an effective manner. The proposed scheme has been evaluated for image permutation analysis and correlation analysis as follows. 

\begin{figure}[!b]
    \centering
    \begin{subfigure}[t]{0.45\linewidth}
        \includegraphics[width=\linewidth]{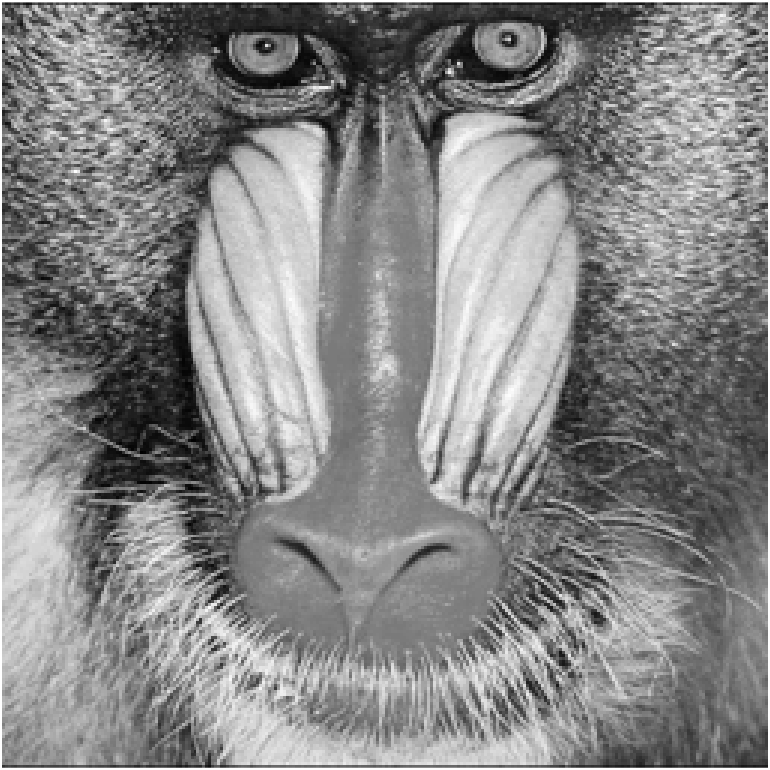}
        \caption{The Original Image}
        \label{fig:fig5-a}
    \end{subfigure}
    \hfill
    \begin{subfigure}[t]{0.48\linewidth}
    \captionsetup{justification=centering} 
        \includegraphics[width=\linewidth]{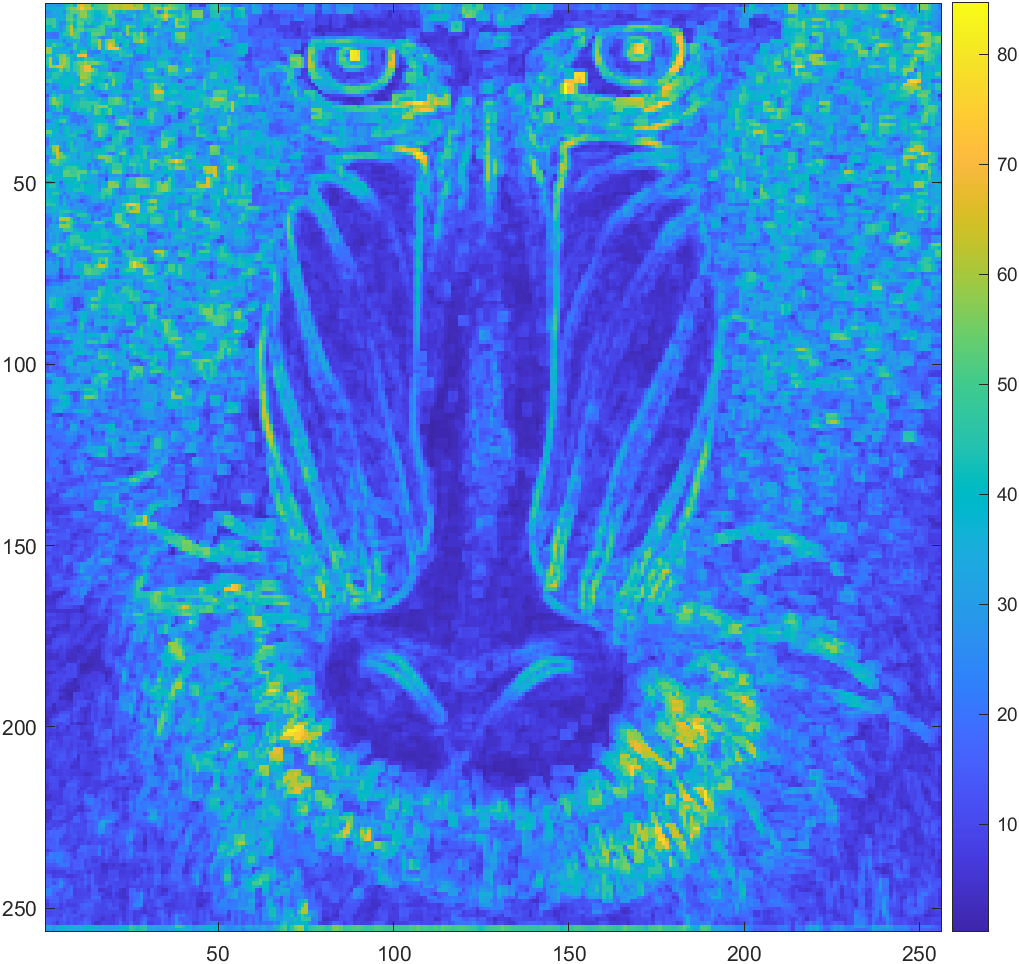}
        \caption{Pixels ranked based on the important feature-information}
        \label{fig:fig5-b}
    \end{subfigure}
    \par\bigskip
    \begin{subfigure}{0.46\linewidth}
        \includegraphics[width=\linewidth]{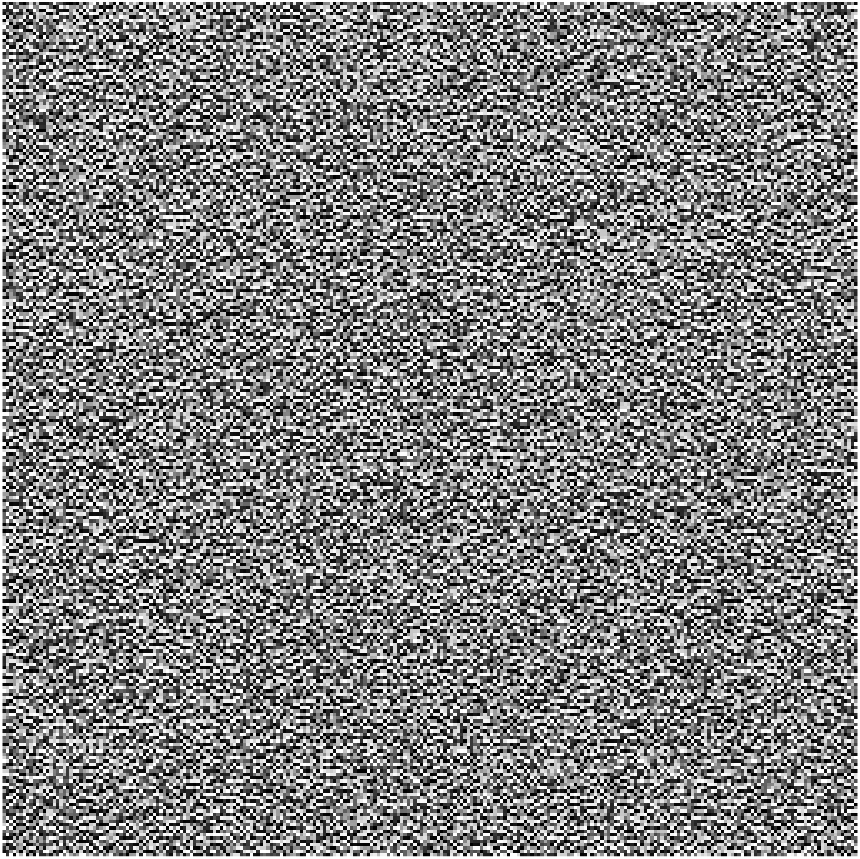}
        \caption{The Permutation Key}
        \label{fig:fig5-c}
    \end{subfigure}
    \hfill
    \begin{subfigure}{0.46\linewidth}
        \includegraphics[width=\linewidth]{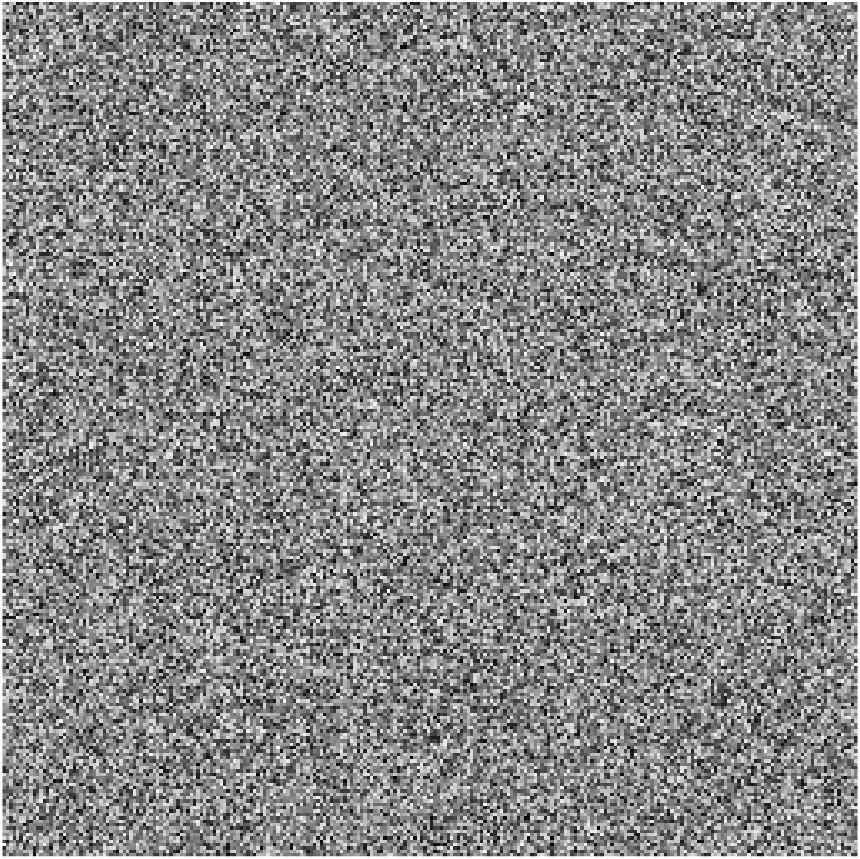}
        \caption{The Final Permuted Image}
        \label{fig:fig5-d}
    \end{subfigure}
    \caption{Permutation Analysis of PermutEx---the proposed diffusion scheme.}
    \label{fig:fig5-permutation_results}
\end{figure}

\begin{figure}[!t]
    \centering
    \begin{subfigure}[t]{0.45\linewidth}
        \includegraphics[width=\linewidth]{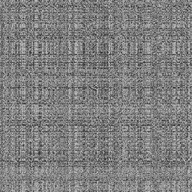}
        \caption{Random row-column shuffling (Correlation=0.0412)}
        \label{fig:fig5-a}
    \end{subfigure}
    \hfill
    \begin{subfigure}[t]{0.45\linewidth}
    \captionsetup{justification=centering} 
        \includegraphics[width=\linewidth]{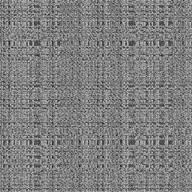}
        \caption{Chaotic row-column shuffling (Correlation=0.0321)}
        \label{fig:fig5-b}
    \end{subfigure}
    \par\bigskip
    \begin{subfigure}{0.45\linewidth}
        \includegraphics[width=\linewidth]{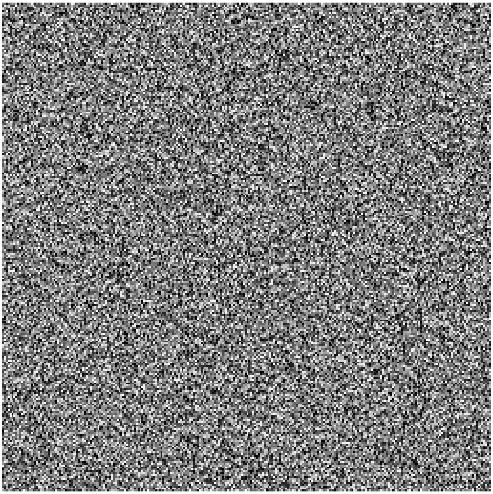}
        \caption{PermutEx without feature-based pixel ranking (Correlation=0.0223)}
        \label{fig:fig5-c}
    \end{subfigure}
    \hfill
    \begin{subfigure}{0.45\linewidth}
        \includegraphics[width=\linewidth]{images/scrambled_results.png}
        \caption{PermutEx with feature-based pixel ranking (Correlation=0.0011)}
        \label{fig:fig5-d}
    \end{subfigure}
    \caption{Comparison of PermutEx with traditional permutation schemes and PermutEX with and without applying feature-extraction-based pixel ranking.}
    \label{fig:fig6-comparison_images}
\end{figure}

\begin{figure}[!ht]
    \centering
    
    \begin{subfigure}{0.45\linewidth}
        \centering
        \includegraphics[width=0.9\linewidth]{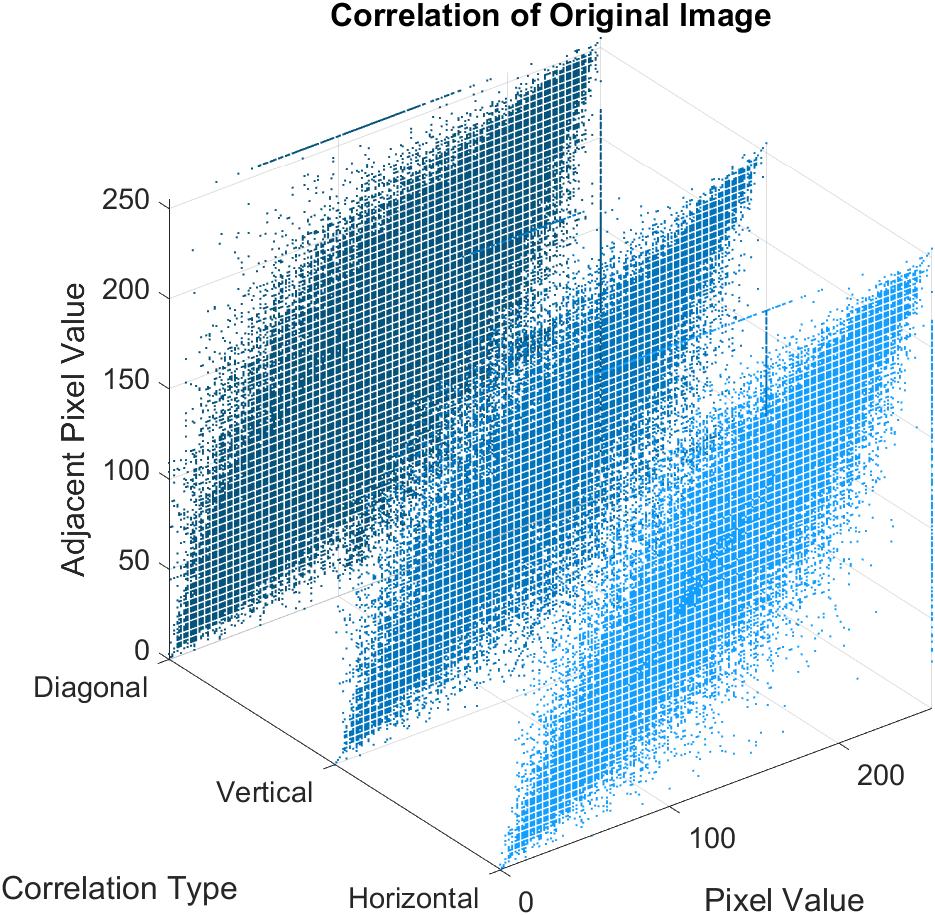}
        \caption*{(a)}
        \label{fig:fig4-a}
    \end{subfigure}
    %
    \begin{subfigure}{0.45\linewidth}
        \centering
        \includegraphics[width=0.9\linewidth]{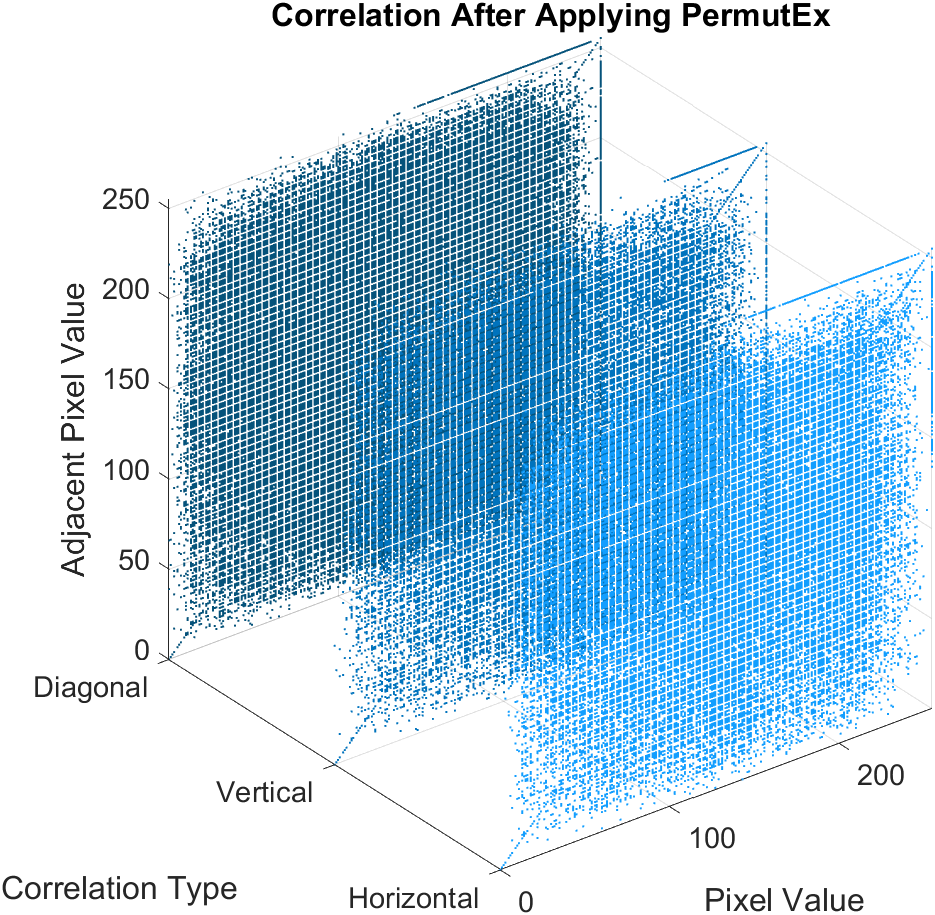}
        \caption*{(b)}
        \label{fig:fig4-b}
    \end{subfigure}
    
    \caption{Correlation analysis of the PermutEx-permuted image}
    \label{fig:fig7-correlation}
\end{figure}

\subsection{Permutation Analysis}
The ranking-based permutation ensures that the most "important" or "information-rich" pixels are treated differently, making it more secure. Fig. \ref{fig:fig5-permutation_results} presents the permutation results of the proposed scheme depicting the the pixels ranked on the basis of extracted information, the unique chaotic permutation key, and the final permuted image. Moreover, the comparison of the proposed permutation scheme with traditional permutation schemes is given in Fig.\ref{fig:fig6-comparison_images}. It also displays the comparison of PermutEx with and without applying the feature-based pixel ranking.

\begin{table*}
    \centering
    \caption{Correlation analysis of the PermutEx permuted image and comparison with other techniques}
    \label{table1}
    \begin{tabular}{|p{3cm}|p{2.5cm}|p{2.5cm}|p{2.5cm}|p{2.5cm}|p{2.5cm}|}
        \hline
        \textbf{Image} & \textbf{Horizontal Coeff} & \textbf{Vertical Coeff} & \textbf{Diagonal Coeff} & \textbf{GLCM Correlation} & \textbf{Correlation (Orig, Permuted) – Corr2} \\
        \hline
        Original Baboon Image & 0.8824 & 0.8397 & 0.7990 & 0.7915 & 1 \\
        \hline
        Random row-column shuffling & 0.4493 & 0.4533 & 0.2009 & 0.0412 & 0.003721 \\
        \hline
        Chaotic row-column shuffling & 0.4405 & 0.4578 & 0.1970 & 0.0321 & 0.028485 \\
        \hline
        PermutEx without feature-based pixel ranking & 0.3265 & 0.3617 & 0.1102 & 0.0223 & 0.037976 \\
        \hline
        \textbf{PermutEx with feature-based pixel ranking} & \textbf{0.0021} & \textbf{-0.0029} & \textbf{0.0022} & \textbf{0.0011} & \textbf{0.000062} \\
        \hline
    \end{tabular}
\end{table*}

\subsection{Correlation Analysis}
Extensive correlation analysis of the proposed scheme has been carried out. Results of the correlation analysis with comparison are given in Table \ref{table1} showing an ideal value of 0.000062 for the PermutEx-permuted images. The correlation between adjacent pixels is displayed in Fig. \ref{fig:fig7-correlation}. The correlation plots are uniformly distributed, which is an excellent indicator of a good permutation scheme as it shows that the pixel dependencies have been well diffused.

\section{Conclusion}
This paper introduced PermutEx, a new technique to scramble pixels effectively. Unlike traditional methods, PermutEx uses the intrinsic image-information-based features, i.e., spatial frequency and local contrast, to scramble the pixels of the plaintext image. This feature-driven permutation leads to effective disruption in correlation of the information-rich areas within the image. This feature-based ranking makes PermutEx more secure and effective in term of diffusing the pixels. PermutEx achieved a very low correlation value of 0.000062 making PermutEx extremely efficient, effective, and secure method of image permutation and is recommended to be used in image encryption algorithms as a diffusion technique.

\bibliographystyle{IEEEtran}
\bibliography{myreferences}

\end{document}